\documentclass[longnamesfirst,aps,prd,twocolumn,groupedaddress,byrevtex,nofootinbib,amsmath,showpacs]{revtex4}

\shortcites{capozziello99}

\usepackage{graphicx}

\allowdisplaybreaks

\newcommand{\figwidth}{8.cm}

\newcommand{\xfigwidth}{8.6cm}

\newcommand{\diff}{\mathrm{d}}
\newcommand{\Diff}{\mathrm{D}}

\newcommand{\rtext}[1]{\hspace{0.4em}\text{#1}}

\newcommand{\ritext}[1]{\hspace{0.4em}\text{#1}}

\newcommand{\sub}[1]{_\text{#1}}

\newcommand{\supp}[1]{^\text{#1}}

\newcommand{\fov}{FOV}

\newcommand{\fod}{FOD}

\newcommand{\vc}[1]{\mathbf{#1}}

\newcommand{\vdopp}{v\sub{Dopp}}

\newcommand{\wc}{w\sub{c}}

\begin{document}

\title{Deflection of light and particles by moving gravitational lenses}

\author{Olaf Wucknitz}
\email[Electronic address: ]{olaf@astro.physik.uni-potsdam.de}
\affiliation{Universit\"at Potsdam, Institut f\"ur Physik, Am Neuen Palais 10,
     14469 Potsdam, Germany}

\author{Ulrich Sperhake}
\email[Electronic address: ]{sperhake@astro.psu.edu}
\affiliation{Pennsylvania State University, Centers for Gravitational Physics
  \& Geometry and for Gravitational Wave Physics, University Park,
  Pennsylvania 16802, USA}  

\date{19 Jan 2004}
\received{6 Mar 2003}
\revised{2 Sep 2003}
\accepted{24 Nov 2003}

\begin{abstract}
Various authors have investigated the problem of light deflection by radially
moving gravitational lenses, but the results presented so far do not appear to 
agree on the expected deflection angles.
Some publications claim a scaling of deflection angles with $1-v$
to first order in the radial lens velocity $v$, while others obtained a scaling
with $1-2v$. In this paper we generalize the calculations for arbitrary lens
velocities and show that the first result is the correct one.
We discuss the seeming inconsistency of relativistic light deflection with the
classical picture of moving test 
particles by generalizing the lens effect to test particles of arbitrary
velocity, including light as a limiting case. We show that the effect of
radial motion of the lens is very different for slowly moving test particles
and light and that a critical test particle velocity exists for which the
motion 
of the lens has no effect on the deflection angle to first order.
An interesting and not immediately intuitive result is obtained in the
limit of a highly relativistic motion of the lens towards the observer,
where the deflection angle of light reduces to zero.
This phenomenon is elucidated in terms of moving refractive
media. Furthermore, we discuss the dragging 
of inertial frames in the field of a moving lens and the corresponding
Lense-Thirring precession, in order to shed more light on the
geometrical effects in the surroundings of a moving mass.
In a second part we discuss the effect of transversal motion on the observed
redshift of lensed sources. We demonstrate how a simple kinematic calculation
explains the effects for arbitrary velocities of the lens and test
particles.
Additionally we include the transversal motion of the source and observer to
show that all three velocities can be combined into an effective relative
transversal velocity similar to the approach used in microlensing studies.
\end{abstract}

\pacs{95.30.Sf, 04.20.Cv, 04.25.Nx, 98.62.Sb}
\maketitle

\section{Introduction}

The subject of gravitational lenses (light deflection by gravitational fields
in the Universe) is a well-established field in modern astrophysical
research (see, e.g., \cite{blandford92,refsdal94,narayan99,sef}).
Applications
include cosmology, dark matter, the large scale 
structure in the Universe, clusters of galaxies, galactic structure, the
structure of the Milky Way, and even the search for extrasolar planets.
Not to be forgotten, light deflection by the Sun's gravitation was the first
test of the then new theory of general relativity
\cite{dyson20}.

In almost all of the studies, the deflecting masses are treated as being at
rest in the cosmological Robertson-Walker metric or (in the limit of
noncosmological lensing) at rest with respect to the source and observer.
This is well justified since most sufficiently massive astronomical lenses
have velocities very small compared to the velocity of light, which
leads only to minor
gravitomagnetic corrections.

In addition to the potential astrophysical relevance, our main motivation to
study 
moving lenses is the desire to understand the fundamental physics governing
the light deflection caused by such systems in an intuitive and possibly
Newtonian interpretation.
This leads us to the comparison of the deflection of light with the
deflection of slowly moving particles and especially the effect any motion of
the lens has on either.

In contrast to most previous work, we do not only consider
effects to first
order in velocity (\fov)
but for the first time
calculate the deflection for arbitrary velocities of 
both lens and deflected particle (including light in the
ultrarelativistic limit).

To our knowledge, the first calculations of the deflection of light
in
gravitational fields were done by Cavendish (see \citet{will88}) and
\citet{soldner1801}.
This early work is based on
the ansatz of Newtonian particles moving with the speed of light.
\citet{einstein11} used the principle of equivalence to avoid the
physically unsound picture of classical particles for the
description of light. The result is
naturally the same.
Only after the formulation of general relativity the result had to be revised
to be \emph{twice} as large as expected from classical
theory \cite{einstein15}. This relativistic result
has been confirmed by numerous
tests with very high accuracy, see, e.g., \cite{robertson91,lebach95}.

Earlier studies in the context of
moving gravitational lenses were undertaken by
several authors.
In 1993 \citet{pyne93} calculated
the effects of the lens' motion on the
deflection in a Minkowski background metric.
For radial motion
with velocity $v$ they found the deflection
angle to scale with a factor $1-v$ in
\fov\ approximation. The
deflection increases if light and lens 
are moving in opposite directions (as seen by the observer).
Interestingly, the leading term is of first order; i.e., the
effect is not
merely a consequence of scaling with the special-relativistic
parameter $\gamma=1/\sqrt{1-v^2}$\,.

In contrast to this, \citet{capozziello99} found a different
scaling behavior of $1-2v$ for point-mass lenses.
They later generalized their
calculations to other mass distributions with the remarkable but questionable
result that the 
scaling of the deflection caused by a rigidly moving lens does not only
depend on the mass distribution and velocity of the lens but also on its
internal parameters \cite{capozziello01}.
This is of particular significance because it would enable us
to derive information from such observations about the inner structure of
lenses beyond their mass and momentum.
The same scaling for point mass lenses was found by \citet{sereno02a} and later
generalized for other mass distributions in \cite{sereno02b}.

An alternative analytic method was used by \citet{frittelli02a} which
confirmed the result of \cite{pyne93}. \citet{frittelli02b} also discusses
the discrepancy between the two results, again favoring \cite{pyne93}.
Below we will demonstrate how we find our results to agree with those of
\cite{pyne93,frittelli02a,frittelli02b,molnar02}, whereas our velocity
corrections differ by a factor of 2 from the values of
\cite{capozziello99,capozziello01,sereno02a,sereno02b}.

A further interesting result of \cite{pyne93} concerns the effect of a
\emph{transversal} motion of the lens
on the observed redshift of the source.
We believe this to be the most
promising possibility of actually measuring gravitomagnetic effects in
gravitational lensing and even utilizing them for astrophysical
studies. Concrete practical
scenarios are discussed by \citet{molnar02}.

\citet{kopeikin99} presented an exhaustive calculation of light
propagation in the field of an ensemble of arbitrarily moving point masses in
terms of retarded Li\'enard-Wiechert potentials, covering effects on both
deflection angles and redshifts.

Aside from gravitational lenses of standard astronomical origin, the
deflection of light by
topological defects in the universe, such as cosmic strings,
has stimulated a great deal of scientific investigation.
Of most interest in the context of our work is
the study of moving cosmic strings
(see, for example, \citet{deLaix1996} and \citet{uzan}).
Below we will
discuss the similarities and differences arising from such different types
of lenses and motions.

This paper is organized as follows. After a summary of the
notation and approximations used in this work,
we begin our discussion with the analysis of purely radial motion
of the gravitational lens in Sec.~\ref{sec:radial}.
In view of the fact that the gravitational
lensing effect is commonly described in the framework of Newtonian
physics, it will be important to verify to what extent this
approximation remains valid for \emph{moving} lenses.
For this purpose we will present a
Newtonian calculation of the deflection of massive test particles
in Sec.~\ref{sec:class}
and contrast the results with the relativistic expressions
derived in Secs.~\ref{sec:obs frame} and \ref{sec:lens frame}.
The dependency of the deflection on the test particle's velocity will
be studied in detail in Sec.~\ref{sec:disc radial}.
We then turn our attention to the effect of transversal motion of the
lens in Sec.~\ref{sec:transverse} and conclude in
Sec.~\ref{sec:conclusions}.

A comparison of our analytic results for radial motion with more
accurate numerical simulations is presented in Appendix~\ref{sec:geoexac}.
An illustration of
how some of the calculated effects can be understood in a more
intuitive fashion in terms of moving refractive media is given in
Appendix~\ref{sec:refrindex}.
Finally we discuss in Appendix~\ref{sec:framedrag} the dragging
of inertial frames in the field of a moving lens by calculating the
Lense-Thirring precession in this case.

\section{Notation and approximations}

In this work we will measure velocities in units of the speed of light,
i.e., we set $c=1$.
The velocity of the lens is denoted by $v$ and that of the test
particle by $w$. The deflection of
light is obtained in the limit $w=1$. Without loss of generality
we perform all calculations in ``2+1'' dimensions, described by
Cartesian coordinates $y$ and $z$ and time $t$.
For all relativistic calculations we will use the metric
signature $(+--)$.

Unless stated otherwise
we use the weak-field and small-angle approximation
as is common in gravitational lens theory. We restrict our discussion to thin
lenses so that it is sufficient to calculate the deflection of light in a
cosmologically small region
around the lens only.
We are thus able to describe the surroundings of the lens
by the Minkowski metric.
The local deflection angle calculated in this way can then be used as part of
the cosmological lens equation in which it is related with cosmological
  distances and the \emph{observed} 
positional displacement,
see, e.g., \cite{sef}.

In the weak-field limit, the line element in the \emph{rest frame of the lens}
can be written as 
\begin{equation}
\diff s^2 = \left(1+2\Phi\right)\diff t'^2 -
\left(1-2\Phi\right)\left(\diff y'^2+\diff
  z'^2\right) 
\rtext{.}
\label{eq:metric static}
\end{equation}
Here we have used primed coordinates to distinguish the rest frame
of the lens from that of the observer (unprimed coordinates).
The static Newtonian
gravitational potential $\Phi$ is linear in the mass distribution and
satisfies  $|\Phi|\ll1$.

The unperturbed light/particle is traveling in the positive $z$ direction at
$y=0$ with velocity $w$ as measured in the unprimed observer system.
The deflected trajectories will therefore be directed
at small angles relative to the $z$ axis.
The only requirement we impose on the velocity of the test particle is
that it travels faster than the lens, 
i.e., $w>v$, so that it passes the lens in the usual direction.

The approximations imply that deviations of
$y$ from
the unperturbed path $y\equiv 0$ need
to be calculated to first order only.
This ``first order in deflection'' (\fod) approximation is not to be confused
with \fov. Below, these approximations will allow us
to integrate along the unperturbed path instead of the deflected (and then
still unknown) path itself. We will further be able to
compare the deflection of light for lenses at different velocities
without changing the light path close
to the lens, i.e., without changing the impact parameter.
In this approximation
we can also relate the coordinates along the
path of the test particle by
\begin{equation}
\diff z = w\,\diff t
\rtext{.}
\label{eq:dz=wdt}
\end{equation}

The deflection angle is the difference of the propagation directions before
and after passing the lens. For small angles and propagation close to the
$z$ direction, this can be written as
\begin{equation}
\alpha = \left.\frac{\diff y}{\diff z}\right|\sub{out}  -
\left.\frac{\diff y}{\diff z}\right|\sub{in}  \rtext{.}
\label{eq:def defl}
\end{equation}
For the ingoing path we have $z\to-\infty$, for the outgoing $z\to+\infty$.

For the radial velocity of the lens and the test particle
we will adopt the sign-convention that
they are positive if directed towards the observer so that $v$ and $w$
are measured in the same direction.
This definition is opposite
to the usual astronomical definition of radial velocities but is commonly used
in the literature on moving gravitational lenses.

\section{Radial motion}
\label{sec:radial}

\subsection{Nonrelativistic calculation based on \\ the principle of
  equivalence\label{sec:class}} 

We begin our discussion of the effect of purely radial motion of the lens
on the resulting deflection angle with a Newtonian discussion.
Aside from providing an intuitive insight into the effect, the
results thus obtained will also enable us below to highlight
the quantitative and qualitative modifications
arising in a general relativistic treatment.

In order to avoid a description of light in terms of classical particles,
we follow the lines of \citet{einstein11} and apply the principle of
equivalence to the gravitational effects
and use nonrelativistic kinematics otherwise.
At each point of the path of a test particle, we can define a freely falling
observer who is momentarily at rest and who
views the deflected path in her vicinity as a straight
line.
The equations of motion of this observer
are given by
\begin{alignat}{2}
\frac{\diff^2 y}{\diff t^2} &= - \Phi_y \rtext{,}  &
\qquad \frac{\diff^2 z}{\diff t^2} &= - \Phi_z \rtext{,}
\label{eq:motion obs}
\end{alignat}
where $\Phi_y$ and $\Phi_z$ denote partial derivatives of the potential
with respect to $y$ and $z$, respectively.
In the frame of this observer, the acceleration of the test particles
vanishes, so that they follow
the same equations of motion \eqref{eq:motion obs} as the falling observer.

In order to calculate the deflection angle defined in Eq.~\eqref{eq:def defl}
we need to calculate $\diff y/\diff z$. In \fod\
this is obtained from Eq.~\eqref{eq:motion obs}
by replacing time derivatives with $z$ derivatives according to
Eq.~\eqref{eq:dz=wdt} and integrating once over $z$.
We thus obtain for the deflection angle
\begin{equation}
\alpha\supp{cl}_w(v) = -\frac{1}{w^2}\int_{-\infty}^\infty\diff z \, \Phi_{y}
 (y,z)\rtext{.}
\end{equation}
Next we transform the integral to primed coordinates.
First we note that the boost does not affect the $y$ direction, so
that $\Phi_y(y,z)=\Phi_{y'}(y',z')$. Secondly we can use
Eq.~\eqref{eq:dz=wdt} together with
the nonrelativistic limit of the Lorentz transformation
\eqref{eq:lotra iii} below to obtain $\diff z'=(1-v/w)\,\diff z$.
This gives us the deflection angle as
\begin{equation}
  \alpha\supp{cl}_w(v) = -\frac{1}{w(w-v)}\int_{-\infty}^\infty\diff
  z' \, \Phi_{y'} (y',z')\rtext{.}
\end{equation}
The integral is now expressed entirely in terms of the coordinates
of the lens' rest frame and is therefore independent of the velocity
$v$ (the perturbation of the gradient $\Phi_{y'}$ on a perturbed path
is of second order in the deflection angle and is thus ignored in
\fod). We can therefore express
the dependency of the deflection angle on the velocity $v$ of the
lens in \fov\ approximation as
\begin{align}
\frac{\alpha\supp{cl}_w(v)}{\alpha\supp{cl}_w(0)} &= \frac{w}{w-v}
\rtext{,}
\label{eq:alpha class i}
\\
\alpha\supp{cl}_w(0) &= - \frac{1}{w^2}\int_{-\infty}^\infty\diff z' \,
\Phi_{y'}\rtext{.}
\label{eq:alpha class ii}
\end{align}
The factor $1/w^2$ in the last equation can be interpreted as follows.
One factor $1/w$ originates from Eqs.~\eqref{eq:dz=wdt}
and \eqref{eq:def defl} 
and represents the geometrical effect that
the transversal acceleration due to
the gradient $\Phi_{y'}$ and the resulting change in velocity along the
$y$ direction
$\Delta w_\bot$ correspond to a larger deflection angle the smaller
the velocity-component $w$ along the $z$ direction.
The second factor $1/w$ is due to the
interaction time which scales inversely with the velocity $w$.
The first contribution is unchanged for moving lenses while the interaction
time now scales inversely with the \emph{relative} velocity
$w-v$, leading to the correction factor in Eq.~\eqref{eq:alpha class i}.

This result is not only exact in the nonrelativistic limit
($w \rightarrow 0$)
but still provides a decent
approximation for the deflection angle of \emph{light} for lenses
  \emph{at rest}:
The
only modification caused by general relativity is an additional factor of 2.
One may ask
whether this
remains true for \emph{moving} lenses
--- can one simply apply the missing scaling factor of 2 to the
deflection angle of light in the moving case as well?
Below we will learn that this is not the case.

As a qualitative result for slowly moving particles,
we find from Eq.~\eqref{eq:alpha class i} that
the deflection increases (due to the change in interaction time)
if the test particle and lens are moving in the same direction.
We will see that the opposite holds for light, so
there exists a test particle velocity $\wc$ for which motion 
of the lens has no effect on the deflection angle in \fov\
approximation. We will study this 
feature in more detail in the next sections when we calculate
the deflection in the framework of general relativity.

\subsection{Relativistic calculation in the observer frame\label{sec:obs frame}}
In order to calculate the trajectories of test particles in the observer's
rest frame, in which the lens is moving, we need to find the line element in
this system.
For this purpose we transform the line element \eqref{eq:metric static}
from the rest frame of the lens (primed) to that of the observer (unprimed
coordinates). The corresponding Lorentz transformation is given by
\begin{align}
\diff t' &= \gamma \left(\diff t- v \, \diff z\right) \rtext{,}
\label{eq:lotra i}
 \\
\diff y' &= \diff y \rtext{,} \\
\diff z' &= \gamma \left(\diff z - v \, \diff t\right) \ritext{,} 
\label{eq:lotra iii}
\end{align}
where $\gamma=1/\sqrt{1-v^2}$.
We drop the \fov\ approximation for this calculation so that the
line element for arbitrary velocity $v$ in the observer's rest frame
is given by
\begin{align}
\diff s^2 = &
\left[1+2\left(1+v^2\right)\gamma^2\Phi\right]\diff t^2 -
  \left(1-2\Phi\right)\,\diff  y^2  \notag \\
& - 
\left[1-2\left(1+v^2\right)\gamma^2\Phi\right]\,\diff
z^2 - 8 v\gamma^2\Phi \,\diff t\, \diff z\rtext{.}
\label{eq:metric moving}
\end{align}
In the limit of small $v$ this metric reduces to the \fov\ result
of \citet{sef}.

The geodesic equation with a given metric is equivalent to an Euler-Lagrange
system of equations with Lagrange function
\begin{equation}
L= \left(\frac{\diff s}{\diff \lambda}\right)^2 \rtext{,}
\label{eq:lagrange}
\end{equation}
where $\lambda$ is an affine parameter of the test particle's world line. We
denote the derivative with respect to this parameter by a dot, e.g.,
$\diff y/\diff \lambda=\dot{y}$.
Using the line element \eqref{eq:metric moving}, the equation for $y$ in \fod\
approximation becomes
\begin{gather}
\frac{\diff}{\diff\lambda} \frac{\partial L}{\partial \dot y} -
\frac{\partial L}{\partial y} = 0  \rtext{,}
\label{eq:elg y}
\\
\ddot y = -\left[ \left(1+v^2\right)\left(\dot t^2+\dot
    z^2\right) - 4v\,\dot t\dot z \right] \gamma^2 
\Phi_y \rtext{.}
\label{eq:ddot y}
\end{gather}
Terms like $\dot{\Phi}\dot y$, $\Phi\ddot{y}$ and ${\dot y}^2$ are neglected
here because they are of higher order in the deflection.

From the corresponding
equations for $z$ and $t$ we learn that deviations from the 
undeflected path in these coordinates are of the same order as the deflection
in $y$ so that they become insignificant after multiplication with $\Phi_y$
in Eq.~\eqref{eq:ddot y}. We can therefore apply
$\dot t=\dot z/w$ from Eq.~\eqref{eq:dz=wdt} for the undeflected path
and use $\lambda=z$ as the affine parameter.
Equation~\eqref{eq:ddot y} can then be written as
\begin{equation}
\frac{\diff^2 y}{\diff z^2} = -\left[ (1+v^2)(1+w^{-2}) -
  4\frac{v}{w} \right] \gamma^2 \Phi_y \label{eq: d2ydz2_rel}
\rtext{.}
\end{equation}
After integration we can use Eq.~\eqref{eq:def defl} and find the deflection
angle 
\begin{align}
\alpha_w (v) &= \int_{-\infty}^\infty \diff z \, 
\frac{\diff^2 y}{\diff z^2} \label{eq: integral_rel} \\
&= \frac{1}{(1-v/w)\gamma} \int_{-\infty}^\infty \diff z' \, \frac{\diff^2
  y}{\diff z^2} \rtext{.}
\label{eq:factor scale 2}
\end{align}
In the final step we applied a parameter transformation from $z$ to $z'$ using
the unperturbed particle path \eqref{eq:dz=wdt} and the Lorentz transformation
\eqref{eq:lotra iii}.
This result can again be written as a correction to the deflection angle
$\alpha_w(0)$ caused by a lens at rest, so that we obtain
\begin{align}
\frac{\alpha_w (v)}{\alpha_w(0)} &=  \frac{\gamma}{1-v/w} \left( 1+v^2 -
  \frac{4vw}{1+w^2} \right) \rtext{,}
\label{eq:alpha c}
\\
\alpha_w (0) &= -(1+w^{-2}) \int_{-\infty}^\infty \diff z'
\, \Phi_{y'}\rtext{.}
\label{eq:alpha 0 c}
\end{align}
As before
the integral in Eq.~\eqref{eq:alpha 0 c} is evaluated
to \fod\ by using the
unperturbed path. We emphasize that in this approximation
the impact parameter, and thus the integral in Eq.~\eqref{eq:alpha 0 c}, is
a background quantity, i.e.\ remains constant for different
values of $v$ and $w$. In all our comparisons of deflection angles obtained
for different velocities we will therefore consider the impact parameter
as being kept constant, even in the numerical calculations below
which go beyond \fod.

For the special case of light and a lens at rest ($w=1$, $v=0$),
the comparison of Eq.~\eqref{eq:alpha 0 c} with the classical
result \eqref{eq:alpha class ii} yields
the additional factor of 2 introduced by general relativity.  
In the nonrelativistic limit $|v|,w\ll1$ we recover the classical result
\eqref{eq:alpha class i}.

\begin{figure*}[t]
  \centering
  \includegraphics[bb=40 30 572 725,angle=-90,width=0.45\textwidth,clip]{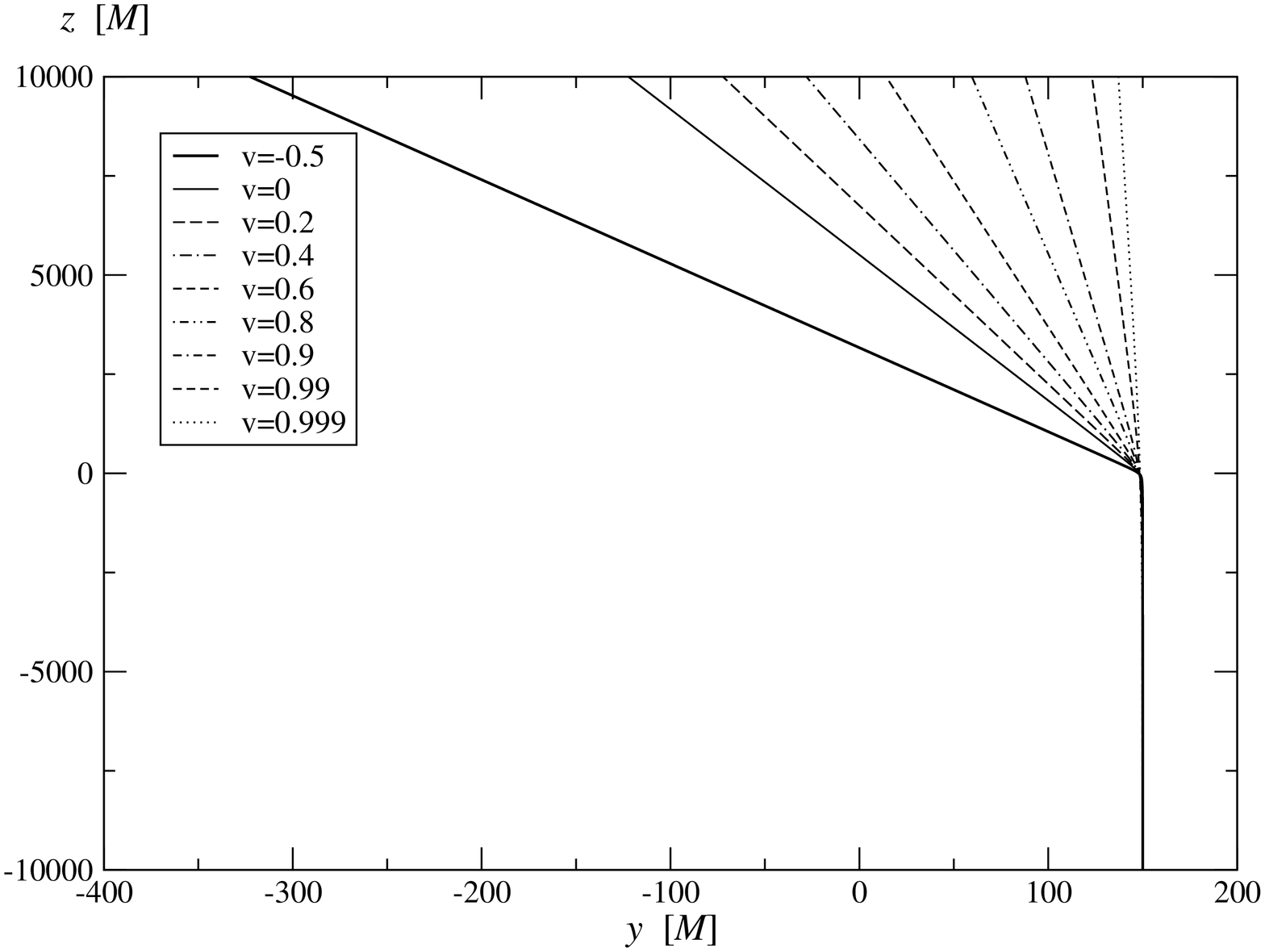}\hspace{0.05\textwidth}%
  \includegraphics[bb=40 30 572 725,angle=-90,width=0.45\textwidth,clip]{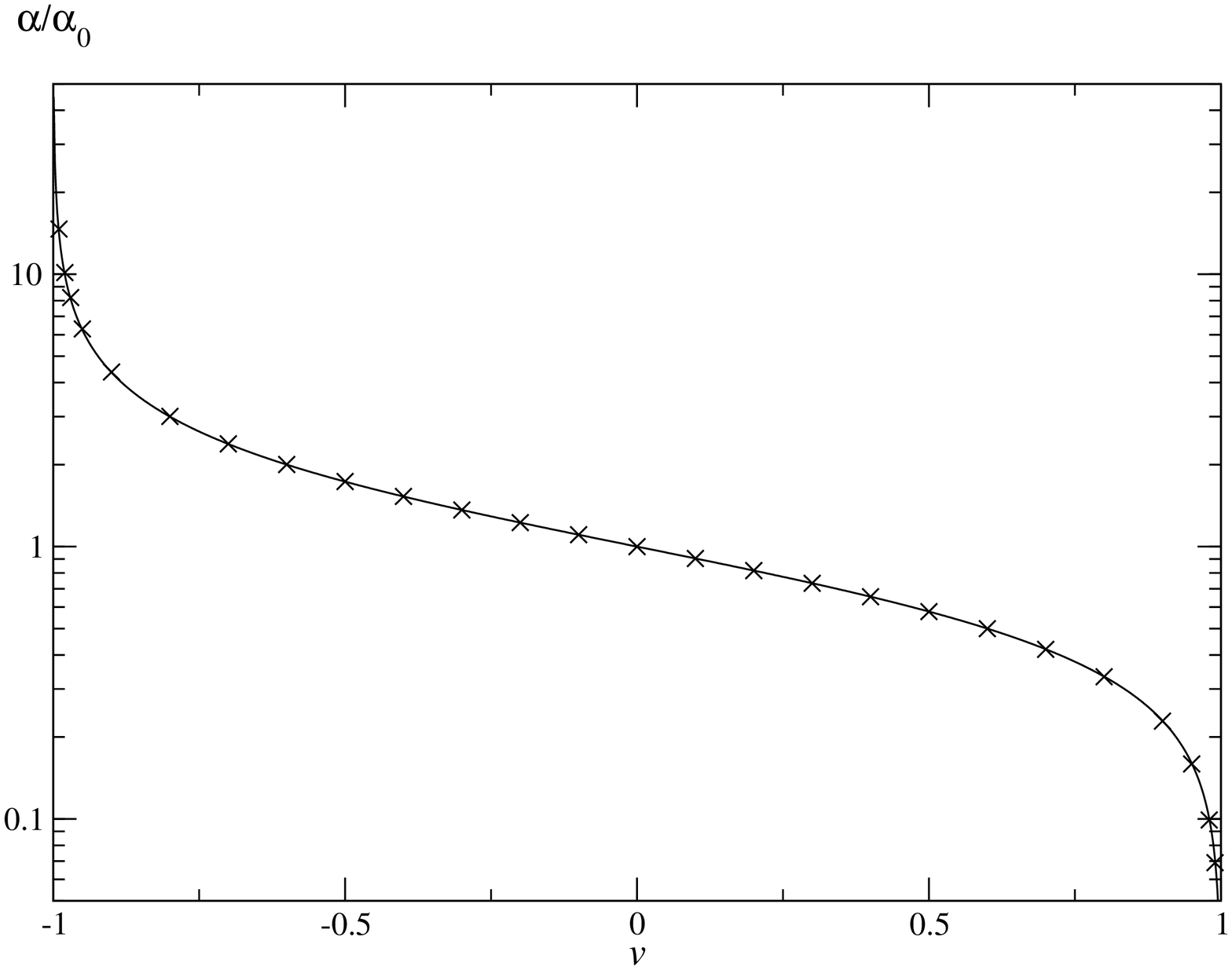}
  \caption{The null-geodesics (left panel) and the (logarithmic) deflection
    angle 
           (right panel)
           are plotted for various velocities of the lens.
           In the right panel the crosses ($\times$) are the
           numerically calculated
           deflection angles, the solid curve
           is given by the analytic result
           \eqref{eq:scaling light} obtained in \fod\ approximation.
           Even though these deflection angles are larger than
           typical astronomical
           values by several orders of magnitude
           (for this calculation $\alpha_0=0.027$), the \fod\ approximation
           still provides excellent results.
           Following the arguments in Sec.~\ref{sec:lens frame} it can be
           shown that, even for $v$ quite close to $-1$, the relative error
           (due to the \fod\ approximation) of $\alpha(v)$ as defined in
           Eq.~\eqref{eq:def defl} is the same as that 
           of $\alpha(0)$ even though the small-angle approximation would not
           hold for $\alpha(v)$ directly.
           }
  \label{fig:deflexac}
\end{figure*}

\subsection{Relativistic calculation in the lens frame\label{sec:lens frame}}

In view of the discrepancy between the results
presented in the literature (see, e.g., \cite{pyne93}
and \cite{capozziello99}), we consider it appropriate to
calculate the same effect again in the lens' reference frame.

For this purpose we consider Eq.~\eqref{eq:alpha 0 c} which
remains valid \emph{in the rest frame of the lens} even for nonzero $v$
and which can be derived directly from the usual static weak-field metric
  \eqref{eq:metric static}.

We must take into account in that case, however, that the
particle velocity now is $w'$ (instead of $w$) and that the resulting
deflection angle will be $\alpha'$, i.e., the angle as viewed in the lens'
system. In order to facilitate a comparison with the results of the
previous section we need to Lorentz-transform both quantities
into the observer's rest frame. We
first consider the velocity $w'$.

Because of the \fod\ approximation we can consider the velocities
$w$, $w'$ as the components along the $z$ axis for this transformation.
This is reflected by Eq.~\eqref{eq:dz=wdt} and its analogue
in primed coordinates $\diff z'=w'\diff t'$. Combining these relations with the
Lorentz transformation \eqref{eq:lotra i}--\eqref{eq:lotra iii}
we directly obtain the relativistic addition of velocities
\begin{equation}
w' = \frac{w-v}{1-vw}\rtext{.}
\label{eq:vel comp}
\end{equation}

We simplify the transformation of the deflection angle $\alpha$ by
defining the direction of the ingoing path as that of zero deflection.
We therefore need to transform the outgoing path only which is parametrized by
\begin{align}
\diff z' &= w' \, \diff t' \rtext{,} \\
\diff y' &= \alpha'_{w'}(0) \, \diff z'
\end{align}
[cf.\ Eqs.~\eqref{eq:dz=wdt} and \eqref{eq:def defl}]. 
Again we use the Lorentz transformation
\eqref{eq:lotra i}--\eqref{eq:lotra iii}
to eliminate the primed coordinate differentials
and Eq.~\eqref{eq:def defl} to express the result in terms of the
deflection angle in the \emph{observer's rest frame}. We thus obtain
\begin{align}
\frac{\alpha_w(v)}{\alpha'_{w'}(0)} = \left(1-\frac{v}{w}\right)\gamma
\rtext{.}  
\label{eq:lorentz contrib}
\end{align}
Eventually we want to compare deflection angles
for \emph{equal test particle
velocities} $w$, so that we still need to express $\alpha'_{w'}(0)$
in terms of $\alpha_w(0)$.
From Eq.~\eqref{eq:alpha 0 c}
we directly obtain
\begin{eqnarray}
\frac{\alpha'_{w'}(0)}{\alpha_w(0)} &=& \frac{1+w'^{-2}}{1+w^{-2}} \nonumber\\
&=& \left(1-\frac{v}{w}\right)^{-2} \left(1+v^2-\frac{4vw}{1+w^2}\right) 
\rtext{,}
\label{eq:w w'}
\end{eqnarray}
which becomes unity in the special case of light ($w=w'=1$).
The combination of Eqs.~\eqref{eq:lorentz contrib} and \eqref{eq:w w'}
confirms the above result \eqref{eq:alpha c} calculated directly from the
viewpoint of the observer.

\subsection{Discussion of radial motion\label{sec:disc radial}}

\subsubsection{Rigidly moving lenses}

We now turn our attention to the interpretation of the deflection of light for
a rigidly moving gravitational lens to \fod.
For this purpose we consider the limit
of Eq.~\eqref{eq:alpha c} for $w\to1$. In that case
the scaling has the simple form
\begin{align}
\frac{\alpha(v)}{\alpha(0)} = (1-v)\,\gamma =\sqrt{\frac{1-v}{1+v}} \rtext{.}
\label{eq:scaling light}
\end{align}
To \fov\ this becomes $1-v$ which confirms the result of
\cite{pyne93} while it differs by a factor of 2 from that given in
\cite{capozziello99}.
We further note that
the quotient $\alpha(v)/\alpha(0)$ diverges for $v\to -1$ but vanishes
in the limit $v\to +1$. Thus, rather surprisingly,
a lens approaching the observer with a highly relativistic velocity will
\emph{not} deflect the passing light at all
even though the effective mass-energy
of the lens as viewed by the observer diverges in this limit.

In view of the counterintuitive nature of this result one may
ask whether the approximation of small deflection angles (FOD) underlying
this calculation is still justified in the limit of $v \rightarrow 1$.
In order to clarify this point we have numerically solved the full
geodesic equations, e.g., Eq.\,\eqref{eq:elg y}, for the case
of a point mass ($\Phi=-M/r$)
\emph{without}
FOD approximation. Instead of Eq.~\eqref{eq:ddot y}
we obtain in this case a system of three nontrivial second order
ordinary differential equations for $t$, $y$, and $z$. These
equations together with more details of the numerical treatment are
listed in Appendix~\ref{sec:geoexac}. Note that the only approximation
in this numerical calculation is the assumption of a weak gravitational
field
in the \emph{rest frame of the lens}. In Fig.~\ref{fig:deflexac} we plot
the resulting geodesics and the
deflection angles obtained for different velocities
of the lens
together with the analytic result in FOD approximation
as given by Eq.~\eqref{eq:scaling light}. For these calculations source
and observer are located at $z= \pm 10\,000\,M$,
respectively, while the impact
parameter, which equals the $y$ position of the source with high accuracy,
is $150\,M$.
The precise values of these parameters do not affect our results.
The lens is always positioned at $y=0$ and moves along the $z$ axis, so
that it reaches $z=0$ at the same time as the null-geodesics.
Note that the $y$ position of the observer depends on the deflection
angle since the impact parameter is kept constant as
the lens' velocity $v$ is varied.

The excellent agreement
between the numerical deflection angle and that predicted by the
approximate FOD result
demonstrates the validity of the FOD approximation in our calculations
even for strongly relativistic velocities $v$.

\begin{figure}
\includegraphics[width=\figwidth]{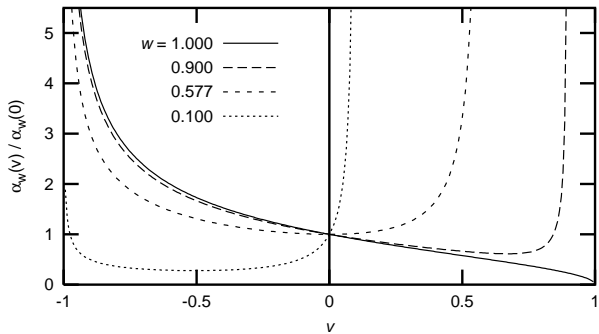}
\caption{\label{fig:v}
  The deflection angle caused by a moving lens
  in units of that arising from a lens at rest as given by
  Eq.~\eqref{eq:alpha c} is shown as a function of $v$.
  The graphs are obtained for $w=1$ (light), $0.9$, $0.577=1/\sqrt{3}$
  ($=w_{\rm c}$), and $0.1$.
  We note that the curve for $w=\wc$ is stationary at $v=0$, i.e., small
  lens 
  velocities have no effect for this critical test particle velocity.
}
\end{figure}
With regard to the deflection of light by cosmic strings
we first note that such types of lenses induce vector and
tensor perturbations of the metric in addition to the scalar
perturbations in our Eq.\,\eqref{eq:metric static}
[cf.\ Eq.\,(38) of \citet{uzan}] which can be combined into an
effective deflecting potential. The deflection angle can be
expressed in terms of this deflection potential and our result
[Eqs.~\eqref{eq: d2ydz2_rel}, \eqref{eq: integral_rel}]
corresponds to the limit of their Eq.\,(45) for
purely scalar perturbations and linear motion of the lens.
In general, \cite{uzan} find the vector and tensor
perturbations to be neglegible as long as the thin lens approximation
is valid but not necessarily for the case of extended lenses.
A time dependence of the deflecting potential which is of particular
interest for string type lenses (because they are expected to move with
relativistic velocities) is that of an oscillating loop.
For this scenario the deflection of light passing sufficiently far
outside the loop has been found to be identical to that of a point mass
source (see \cite{deLaix1996,uzan} for details). In
contrast, the linear motion studied in this work gives rise to a
vector perturbation of the metric, which manifests itself in the
$\diff t\,\diff z$ cross term of the boosted line element \eqref{eq:metric
  moving}. 
The analogy between this term and rotational effects will be studied in
more detail in Appendix\,\ref{sec:framedrag}.

It remains therefore to obtain a better understanding of how
the deflection vanishes in the limit of a lens moving towards the observer
at highly relativistic velocities. For this purpose we
provide in Appendixes \ref{sec:refrindex} and \ref{sec:framedrag} a description
in terms of refractive media and the dragging of inertial frames
similar to the frame dragging close to rotating bodies.

We now return to the general behavior of the deflection angle for
arbitrary velocities $v$ and $w$ which is shown in Fig.~\ref{fig:v}.
This graphic demonstrates the singular behavior of the deflection angle
at
$v=-1$ and (provided $w\ne 1$) $v=w$.
The singularity at $-1$ arises from
the diverging factor $\gamma$ while the second singularity is
a consequence of the vanishing relative velocities
which causes the interaction time to become infinitely large.

\begin{figure}
\includegraphics[width=\figwidth]{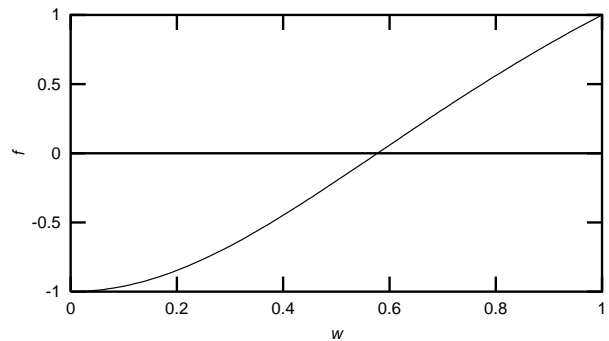}
\caption{\label{fig:c}{Effect of small lens velocities $v$ on the deflection
  angle. According to Eq.~\eqref{eq:scaling slow} the deflection scales
  with $1-(v/w)\,f$, where} $f=(3w^2-1)/(1+w^2)$.
  The graph shows $f$ as a function of the test particle 
  velocity $w$. The effect has opposite signs for $w\to 
  0$ and $w=1$ and there is a critical particle velocity $\wc=1/\sqrt 3$ for
  which the lens motion has no effect to \fov.}
\end{figure}

We next discuss the limit of small lens velocities $v$ (\fov)
combined with a comparatively larger, but otherwise arbitrary
speed of the test particle, i.e., $w \gg v$. In this case the
scaling of Eq.~\eqref{eq:alpha c} simplifies to
\begin{equation}
\frac{\alpha_w(v)}{\alpha_w(0)} = 1- \frac{3w^2-1}{1+w^2} \, \frac{v}{w}
\rtext{.} 
\label{eq:scaling slow}
\end{equation}
We have already mentioned above that the deflection of light
scales with $1-v$ to \fov\ whereas
the Newtonian result of Eq.~\eqref{eq:alpha class i}
predicts a scaling of $1+v/w$ for nonrelativistic test particles
satisfying $w \gg v$.
Hence there must be a critical velocity $\wc$ of the test particle
for which the deflection angle is independent of
the lens' velocity $v$ to \fov. From
Eq.~\eqref{eq:scaling slow} we see that this critical
particle velocity is $\wc=1/\sqrt 3$. We illustrate the functional behavior of
Eq.~\eqref{eq:scaling slow} in Fig.~\ref{fig:c}.

Only after finishing our calculations we learned that special properties of
the velocity $1/\sqrt3$ have been discussed before by Carmeli
[\cite{carmeli72} and problem 5.5.1 in \cite{carmeli82}]. There it was found
that the coordinate 
velocity of a massive particle approaching the center of a Schwarzschild
metric in a radial direction \emph{increases} only for asymptotic starting
velocities 
smaller than $1/\sqrt3$ but \emph{decreases} otherwise, all in a \fod\
approximation. The same was later independently rediscovered by
\citet{blinnikov01,blinnikov03}. It can indeed be shown easily that the same
is true for any 1+1 dimensional weak field with a metric of $\diff
s^2=(1+2\Phi) \diff t^2 - (1-2\Phi)\diff r^2$. To first order in $\Phi$, there
is no coordinate acceleration for $w=\wc$. This effect can in principle be
detected by measuring round trip travel times.

With regard to the observation of the deflection
caused by a radially moving lens we note that
it is not possible to distinguish the scaling of the
deflection angle with $\alpha_w(v)/\alpha_w(0)$ from a scaling with the
total mass of the lens. 
For a given test particle,
the effect can therefore be used to infer a radial motion of the lens
only if an accurate and independent
estimate of the lens' mass is available.
An alternative way to disentangle the two effects
would consist of measuring
deflection angles of test particles with \emph{different}
velocities $w$.
Unfortunately, appropriate test particles other than light are currently not
available in astrophysics. Even if we could measure the direction of
cosmic rays consisting of massive particles with higher accuracy than
is possible with existing instruments, they would not serve our purpose as
they travel at highly relativistic speeds
and would be subject to practically
the same deflection as light.
Furthermore they are also deflected by
electric and magnetic fields.

\subsubsection{Compound lenses}

It is an interesting question whether the gravitomagnetic correction to the
deflection of light or particles could be used to 
deduce information about the inner structure (e.g., the velocity field) of a
lens 
as suggested by the results of \citet{capozziello01} for the deflection
  of light.
We address this question by considering
a compact (relative to the impact parameter $r$) 
lens which we treat as a collection of point masses located close to each
other. 
In terms of our formalism, the argument in favor of the possibility to
probe the inner structure of the lens can be summarized as follows.
Consider a lens with two components of equal mass $m$ moving in
opposite directions with the same speed. Compared to a lens at rest with
mass $2m$, we expect a deflection angle larger by a factor $\gamma$ because
the compound lens will appear as a lens of mass
$2m\gamma$ in the observer's rest frame.
Because of the more complicated dependence of
its right-hand side on $v$, however, Eq.~\eqref{eq:alpha c}
will predict a different deflection angle for this scenario.
It appears therefore that
the deflection is not uniquely determined by the
total four-momentum of the lens.

This argument is not valid, however,
because the compound lens must
be compact not only momentarily but for the whole interval during which the
test 
particle passes by and is deflected significantly.
Otherwise the lens would effectively be extended so that the deflection
would test rather the mass distribution than the internal velocity field
of the lens. Measurable effects are not surprising in that case.

The time scale for the passage is $r/w$
and the displacements of the lens' components during this
interval are $r v/w$. Because these displacements must be $\ll r$
to preserve the compactness of the lens, we obtain an
upper limit for the velocities of its components, 
$|v|\ll w$. In this limit, however, the right-hand side
of Eq.~\eqref{eq:alpha c} becomes approximately
$\gamma (1 + \mathrm{const}\,\, v)$
and the total deflection is indeed determined by the total four-momentum.
We thus conclude that no information about the internal structure beyond
the mass
and total momentum of a compact lens can be obtained from the lensing effect.
In the case of light ($w \rightarrow 1$) this is also demonstrated by
Eq.~\eqref{eq:scaling light}, where the deflection is always determined by
the total
four-momentum, $\alpha\propto p_t - p_z = \sum_j (1-v_j)\gamma_j m_j$.

\subsection{Comparison with previous results\label{sec:compar}}

We will now address the question of what causes the discrepancies between our
results and those in
\cite{capozziello99,sereno02a} where the velocity effects appear to be
overestimated by a factor of 2.
Note that this also affects the ensuing results published in
\cite{capozziello01,sereno02b} which generalize to mass distributions other
than point masses.
Both these groups exclusively use light as test
particles, work in a \fov\ approximation and use essentially the same
approach as \citet{sef} in formulating the problem.\footnote{An alternative
  derivation based on integration of the geodesic
  equation is presented additionally in \cite{capozziello99}. Our arguments
  are valid for this approach as well.}
Following this approach, we use the equivalent formulation of light
propagation in an Euclidean metric with an effective index of
refraction $n$ (cf.\ Appendix \ref{sec:refrindex}).
For a rigidly moving lens with gravitational potential $\Phi$, this index
can be calculated from setting $\theta = 0$ in
Eqs.~\eqref{eq:n'2} and \eqref{eq:n n'} and is given by
\begin{equation}
n-1=-2\,(1-2v)\Phi\rtext{,}
\end{equation}
which does indeed contain the factor $1-2v$.
In order to obtain the correct scaling, however, we must further take into
account that
the motion of the lens
affects the potential $\Phi$ which now becomes
time-dependent in the observer's rest frame.
It is most convenient to calculate the total deflection integral
over the path of
light in the primed coordinates of the 
lens' rest frame.
We thus replace $\Phi(t,y,z) = \Phi(y',z')$ with $z'=z-vt$ from
Eq.~\eqref{eq:lotra iii}
and substitute $z'$ for $z$ in the integral.
The same procedure has been used in Eq.~\eqref{eq:alpha class i} for
the classical
and in Eq.~\eqref{eq:factor scale 2}
for the relativistic calculation.
This correction describes the scaling of time of
interaction caused by the different relative velocity.
The resulting additional factor $1+v$ was \emph{not}
applied in \cite{capozziello99} and
\cite{sereno02a} but would otherwise have changed their scaling factor $1-2v$
to the correct result of a $1-v$ scaling of light deflection in \fov\
approximation.  
\citet{frittelli02b}, on the other hand, \emph{did} apply this factor in her
Eqs.~(33) and (34) and arrives at the same result as we do in
Secs.~\ref{sec:obs frame} and \ref{sec:lens frame} above.

\section{Transversal motion\label{sec:transverse}}

\subsection{Deflection angle}

Until now we have restricted our discussion to a radial motion
of the lens, because this velocity 
component is the only one affecting the deflection angle to \fov.
In this section we will discuss the effects of transversal motion.
For this purpose we use the same approximations as in the discussion of radial
velocity effects, i.e.\ we allow for weak deflections only and work in the
\fod\ approximation.
For radially moving lenses,
the light path across the lens does not depend
on the velocity so that a nonzero
velocity merely gives rise to an overall scaling factor.
This is no longer the case for large transversal velocities.
Due to aberration, the
direction of light propagation in the lens' rest frame \emph{does}
depend on the velocity so that an additional 
term involving the radial derivative $\Phi_{z'}$ appears in the
geodesic equation.
The resulting scaling law
would therefore depend on the mass distribution and is not of general
validity and immediate interest. 

By using the methods of
Secs.~\ref{sec:obs   frame} and \ref{sec:lens frame}, however,
it can be shown
that to \fov\ the deflection
angle is \emph{not} modified at all by transversal motion.

\subsection{Redshift}

Even though a transversal motion of the lens does not affect
the deflection angle to \fov,
there is an interesting effect on the observed redshift of the source.
This can be illustrated as follows.
When the test particle
approaches the lens it falls into the potential well and gains energy. During
the passage, the lens is moving, so that
the potential experienced by the particle as it climbs out of the
potential well will differ from that before. The associated loss in energy
of the test particle will therefore not exactly compensate for its earlier
gain.
The resulting net change in energy can be measured as a
change in the velocity of the test particle or the energy, i.e.,
wavelength, in the case of a photon.

Alternatively one can view the effect in the following way. The test particle
traveling on the rear side of the lens (with respect to the transversal motion)
will be deflected towards the lens and slow down the motion of the lens so
as to preserve total momentum. The energy loss of the lens (which is
proportional to its 
initial velocity, see below) has to be transferred to the particle. For
particles passing on 
the other side of the lens, this effect has the opposite sign.
We will now quantify these ideas straightforwardly by calculating the effect
in the framework of special relativistic kinematics.

The transversal momentum transferred from the lens to the test particle in the
course of the deflection is given by 
\begin{equation}
\Delta p_\bot = \alpha\, p_\|\rtext{,}
\label{eq:alpha delta p}
\end{equation}
where $p_\|$ is the radial momentum of the particle.
We can calculate the accompanying transfer of energy $\Delta E$
by using the fundamental equations of special-relativistic kinematics
\begin{alignat}{3}
E^2&=m^2+p^2 \rtext{,}\qquad &
\vc p&=E \vc v \rtext{,}\qquad &
E&= m\gamma \rtext{.}
\label{eq:kinem}
\end{alignat}
From the first of these equations we derive $E\Delta E=\vc p\cdot \Delta \vc
p$ to first order in $\Delta E$
and thus, together with the second one,
\begin{equation}
\Delta E=\vc v  \cdot\Delta \vc p
\rtext{.}
\label{eq:kinem2}
\end{equation}
We now apply this equation both to the lens and the test particle to write the
equation of energy conservation as
\begin{equation}
v_\|\,\Delta p_\| + v_\bot \, \Delta p_\bot = w\,\Delta p_\|
\rtext{.} \label{eq:energy balance}
\end{equation}
The left-hand side is the energy loss of the lens, written in terms of a
possible radial velocity component of the lens $v_\|$
and its transversal velocity $v_\bot$.
On the right-hand side we have the energy gain of the test particle.
Here the transversal part is of higher order in the deflection and can
be neglected.
Combining Eqs.~\eqref{eq:alpha delta p} and \eqref{eq:energy balance}
we directly obtain the relative change in radial momentum of the particle
\begin{equation}
\frac{\Delta p_\|}{p_\|} = \alpha \frac{v_\bot}{w-v_\|}
\rtext{.}
\label{eq:redshift lens}
\end{equation}
Even though this relation is valid for arbitrary values of the
velocities, we will assume $v_{\|}=0$ for the following discussion.

\begin{figure}
\includegraphics[width=\xfigwidth]{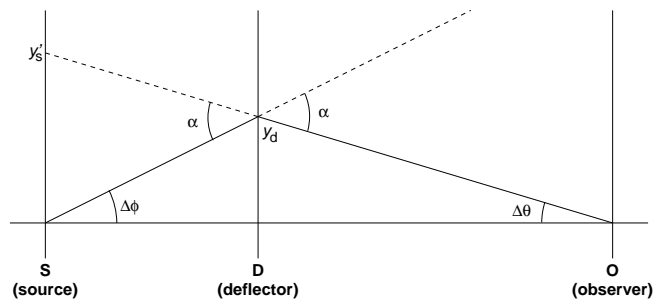}
\caption{\label{fig:geom}The lensing geometry. The solid line shows the
  deflected light path while the horizontal axis is the undeflected light path
  from the source to the observer. The deflection angle is $\alpha$
  and the apparent position of the source is $\Delta\theta$. The
  true position would be at an angle of zero. As shown in the figure, $\alpha$
  is negative while $\Delta\theta$, $\Delta\phi$, and $y\sub d$, $y'\sub s$
  are positive.} 
\end{figure}

In the case of light and assuming $v_\|=0$
we can relate the relative change in momentum to a
shift in frequency or an additional redshift of
$\Delta z=-\alpha\, v_\bot$, which agrees
with the \fov\ results of \citet{pyne93} and \citet{kopeikin99}.
In spite of the remarkably simple derivation, however,
our results remain valid for arbitrary lens potentials and lens and particle
velocities.
Equations~\eqref{eq:alpha delta p}--\eqref{eq:redshift lens} do not even
depend on the physical process which causes the deflection but instead are
valid for any kind of scattering by a massive moving target.

\subsection{Geometry}

So far we have
only considered a transversal motion \emph{of the lens} but assumed that
source and observer are at rest. In reality, source, lens, and observer will
all be moving relative to the cosmological Robertson-Walker background metric.
It is therefore necessary to also
take into account the motion of source and
observer. We will calculate the resulting total redshift
in the framework of the \fov\ approximation.
The fundamental geometry is shown in Fig.~\ref{fig:geom}.
If the source is moving with a transversal velocity $v\sub s$,
this will lead
to a velocity component in the direction of the real path of light
and, thus, a Doppler velocity of $v\sub s \Delta\phi$.
We have already seen in the discussion of
Eq.~\eqref{eq:redshift lens} that a transverse velocity $v\sub d$ of the lens
leads to a  Doppler shift analogous to an apparent
radial velocity $v\sub d \alpha$ of the source. The
motion of the
observer themself gives rise to
an additional contribution corresponding to a
radial velocity of $v\sub o \Delta \theta$.
In \fov\ approximation, the net Doppler
effect is obtained by linear addition
of these velocities because $(1+\Delta z)$ acts multiplicatively, see
Eq.~\eqref{eq:zs' vdopp} below.

If we denote the apparent size distance parameter\footnote{The apparent size
  distance $D$ is defined as the relation between the physical size of a
  distant 
  object $l$ and the (small) angle $\theta$ under which it is seen by an
  observer: $D=l/\theta$}   from $a$ to $b$ (angle 
measured at $a$, size measured at $b$) with $D_{ab}$, we can use two equations
each to express the apparent position of the source $y'\sub s$ and the
position of the lens $y\sub d$ using different angles and distances
(cf.\,Fig.~\ref{fig:geom})
\begin{alignat}{2}
y'\sub s &= \quad - \alpha \, D\sub{ds} &\quad&= \quad \Delta\theta \,
D\sub{os} \rtext{,} \\ 
y\sub d &=  \quad \Delta\theta \, D\sub{od} &\quad&= \quad \Delta\phi \,
D\sub{sd}.
\end{alignat}
These equations can be used to express $\alpha$ and $\Delta\phi$ in terms of
the
apparent displacement $\Delta\theta$ (also called apparent deflection angle)
to write the total Doppler velocity in terms of this angle.
Because of the expansion of the Universe, distance parameters depend on the
direction, i.e., $D\sub{sd}\ne D\sub{ds}$. The two
can easily be related, however, using the redshifts $z\sub s$ and $z\sub d$
of source and lens so that we obtain
\begin{equation}
\frac{D\sub{sd}}{D\sub{ds}} = \frac{1+z\sub s}{1+z\sub d} \rtext{.}
\end{equation}
This leads to the final expression for the total Doppler velocity
\begin{align}
\vdopp = \left( \frac{1+z\sub d}{1+z\sub s} \frac{D\sub{od}}{D\sub{ds}}
  v\sub s -\frac{D\sub{os}}{D\sub{ds}} v\sub d + v\sub o \right)
\Delta\theta
\rtext{.}
\label{eq:v eff}
\end{align}
Remember that our sign convention for $\vdopp$ is opposite to that
commonly used for astronomical
radial velocities, so that the observed redshift $z'\sub s$ is given as a
modification of the unperturbed redshift $z\sub s$ in the following way:
\begin{equation}
1+z'\sub s = (1+z\sub s) (1+\Delta z) = (1+z\sub s)(1-\vdopp)
\label{eq:zs' vdopp}
\end{equation}
In the case of a surrounding Minkowski metric, the redshift factors become
unity and the distance parameters become additive. In this case
Eq.~\eqref{eq:v eff} predicts a zero Doppler velocity if source, lens and
observer are all moving with the same velocity, as expected.
The combination of velocities is very similar to the effective transversal
velocity $V$ 
defined by \citet*[Eq.~(B9)]{krs86} to discuss effects of microlensing. 
The only difference (besides the multiplication with $\Delta\theta$) is that
their $V$ is projected into the source plane and
incorporates a cosmological time dilation factor, so that
\begin{equation}
\vdopp = (1+z\sub d) \frac{D\sub{od}}{D\sub{ds}} \, V  \Delta\theta 
\rtext{.}
\end{equation}

If more than one lensed image of the source is observed,
the effect of the transversal motion can be separated from the {\it a priori}
unknown 
radial velocity (including cosmological redshift) of the
source, which usually contributes a much larger Doppler effect.
In this case the radial velocity will
have the same effect on all images while the effect of transversal motion 
depends on the position of the image. The latter can therefore be measured
unambiguously by comparing the Doppler shifts of at least two images.
This possibility makes the situation conceptually very different from the
effects of radial motion on deflection angles which cannot be
distinguished from a scaling of the total mass of the lens.

Unfortunately the effects are so small that it is hardly possible to
detect them with current state-of-the-art instruments, see e.g.,
\cite{molnar02}. However, telescopes planned for the future, in particular
radio arrays
like ALMA (``Atacama Large Millimeter Array'') or SKA (``Square Kilometer
Array''), will provide sufficient sensitivity and
spectral resolution to measure the effect for sources with
sharp spectral features to facilitate a sufficiently accurate
determination of redshifts.

\section{Summary}
\label{sec:conclusions}

In this work
we used different methods to calculate the effect of \emph{radially} moving
lenses on the deflection of light as well as test
particles with arbitrary velocity. These
calculations are valid for arbitrary
lens velocities and thus generalize the results of previous
publications (e.g., \cite{pyne93,frittelli02a,frittelli02b,molnar02})
dealing with the deflection of light by slowly moving lenses (FOV).
At the same time we have demonstrated how neglecting the time dependence
of
the potential of a moving lens (which is
equivalent to the scaling of the interaction time)
accounts for the discrepancy between our
results and those of \cite{capozziello99,sereno02a,capozziello01,sereno02b}.
In agreement with \cite{frittelli02b} we find
the inclusion of this time
dependence essential for obtaining correct deflection angles.

Our study of the deflection of light and particles by radially moving
lenses has revealed some rather surprising results.
While a motion of the lens parallel to the test particle velocity
increases the deflection of slowly moving particles, as is intuitively
expected from the decreasing relative velocity and increasing interaction time,
the opposite is true for the deflection of light. As a consequence
there exists a critical particle velocity of $\wc=1/\sqrt 3$ for which the
motion of the lens does not have any effect on the deflection to first order.
These results imply that the classical approach of treating light as
classical particles, while providing qualitatively correct results for
lenses at rest (see e.g., \citet{soldner1801}),
does not facilitate a satisfactory description for the
case of radially moving lenses.
This behavior seems to be related to the effects found by
\citet{carmeli72,carmeli82} and \citeauthor{blinnikov01}
\cite{blinnikov01,blinnikov03} for radial motion  
in the Schwarzschild metric where massive particles are accelerated when
approaching the center only for $w<\wc$ but decelerated otherwise.

As a further surprising result
we find that the deflection angle of light vanishes in the limit
of gravitational lenses approaching the observer with
highly relativistic velocities,
even though the effective mass-energy of the lens 
diverges in this limit. Intuitively one would have rather expected a diverging
deflection angle in this scenario.
In Appendix~\ref{sec:refrindex} we provide a descriptive explanation of
this effect by demonstrating 
how gravitational lensing can be formulated 
in terms of light propagating in a refractive medium, both for
static \emph{and}
moving lenses.
In this picture the coordinate velocity of light is reduced in the
rest frame of the lens. For the case of a relativistically moving
lens, however, this reduction is effectively compensated by
the motion of the rest frame of the lens with respect to the
observer, so that the net reduction vanishes.
The unchanged velocity of light in turn implies a vanishing deflection angle.

We have addressed the question whether
the deflection of test particles or light can provide
information about the internal structure of a rigidly moving, compact
lens beyond its mass and total momentum. We find
that this is not possible in the case of lenses that remain compact during
the passage of the test particle.

In Appendix~\ref{sec:framedrag} we additionally extend our discussion
beyond the pure deflection of test particles and calculate the effects of a
moving lens on the inertial system of an observer at rest.
We find the motion of the lens to give rise to a
dragging of this local inertial frame similar to the
frame dragging known in the case of rotating stars or Kerr black holes.

In order to study the effects caused by \emph{transversally}
moving lenses we have used kinematic arguments in the framework of
special relativity.
These calculations predict
a change of the radial momentum of the deflected test particle.
In the case of light this
manifests itself in a change of wavelength. In spite of their simplicity
our kinematic
calculations extend previous discussions (using general relativistic
  integrations along the geodesics) to arbitrary
velocities of the lens and the test particles.

Finally, we have investigated the total Doppler shift arising from
a moving lens in combination with a transversal motion of the source
and the observer.
The resulting effective Doppler
velocity is similar to the total effective transversal velocities as commonly
defined in studies of microlensing effects in cosmological lenses. We have
thus been able to directly relate observable quantities with transversal
velocities of source, lens, and observer. The required accuracy
to apply this method to measure transversal velocities of astronomical
objects, especially galaxies, is beyond
the capability of current astronomical instruments but will
be reached with the next generation of telescopes.

\begin{acknowledgments}
O.W. was supported by the BMBF/DLR Verbundforschung under grant 50\,OR\,0208.
U.S. acknowledges the support of the Center for Gravitational Wave
Physics funded by the National Science Foundation under Cooperative Agreement
PHY-0114375. Work partially supported by NSF grant PHY-9800973 to Pennsylvania
State University.
\end{acknowledgments}

\appendix

\section{The geodesic equations without FOD approximation}
%
%
\label{sec:geoexac}

We consider the boosted weak-field limit of the Schwarz\-schild metric
Eq.~\eqref{eq:metric moving}. The geodesic equations are derived from the
Euler-Lagrange equation \eqref{eq:elg y} and the analogues for $t$ and $z$.
In contrast to the analytic treatment of Sec.~\ref{sec:obs frame}
we do not use the FOD approximation and obtain the equations
\begin{widetext}
\begin{align}
  \begin{split}
  0 =& \left(2 + 4\frac{1+v^2}{1-v^2} \Phi\right) \ddot{t} -
       8\frac{v}{1-v^2} \Phi \ddot{z} -2\frac{v(1+v^2)}{(1-v^2)^{3/2}}
       \Phi_{z'} \dot{t}^2 + 4\frac{1+v^2}{1-v^2} \Phi_{y'}
       \dot{t} \dot{y} \\
     & + 4\frac{1+v^2}{(1-v^2)^{3/2}} \Phi_{z'} \dot{t} \dot{z}
       + 2\frac{v}{\sqrt{1-v^2}}\Phi_{z'} \dot{y}^2
       - 8\frac{v}{1-v^2} \Phi_{y'} \dot{y} \dot{z}
       - 2\frac{v(3-v^2)}{(1-v^2)^{3/2}} \Phi_{z'} \dot{z}^2\rtext{,} \label{eq:GEOT}
  \end{split} \\
  \begin{split}
  0 =& \left(-2+4\Phi \right) \ddot{y}
       - 2\frac{1+v^2}{1-v^2}\Phi_{y'} \dot{t}^2
       - 4\frac{v}{\sqrt{1-v^2}}\Phi_{z'} \dot{t}\dot{y}
       + 8\frac{v}{1-v^2} \Phi_{y'} \dot{t}\dot{z} \\
     & + 2\Phi_{y'} \dot{y}^2
       + 4\frac{1}{\sqrt{1-v^2}}\Phi_{z'} \dot{y}\dot{z}
       - 2\frac{1+v^2}{1-v^2} \Phi_{y'} \dot{z}^2\rtext{,}
  \end{split} \\
  \begin{split}
  0 =& -8\frac{v}{1-v^2} \Phi \ddot{t}
       +\left( -2+4\frac{1+v^2}{1-v^2} \Phi \right) \ddot{z}
       - 2\frac{1-3v^2}{(1-v^2)^{3/2}}\Phi_{z'} \dot{t}^2
       - 8\frac{v}{1-v^2} \Phi_{y'} \dot{t} \dot{y} \\
     & - 4 \frac{v(1+v^2)}{(1-v^2)^{3/2}} \Phi_{z'} \dot{t} \dot{z}
       - 2\frac{1}{\sqrt{1-v^2}}\Phi_{z'} \dot{y}^2
       + 4\frac{1+v^2}{1-v^2}\Phi_{y'} \dot{y}\dot{z}
       + 2\frac{1+v^2}{(1-v^2)^{3/2}} \Phi_{z'} \dot{z}^2\rtext{.} \label{eq:GEOZ}
  \end{split}
\end{align}
\end{widetext}
In order to complete this set of equations
we still need one further condition
which is given by the requirement that the geodesics be null:
\begin{align}
  0 &= g_{tt}\dot{t}^2 + g_{yy} \dot{y}^2 + g_{zz}\dot{z}^2 + 2g_{tz}
       \dot{t} \dot{z} \rtext{.} 
\end{align}

The potential of the lens is assumed to be that of a point mass given in
the weak-field limit by
\begin{align}
  \Phi(t', y', z') &= -\frac{M}{\sqrt{y'^2+z'^2}}
\end{align}
in the lens' rest frame. The resulting potential in the observer's frame
is obtained from Lorentz transformation. For the derivatives that implies
\begin{align}
  \Phi_{y} &= \Phi_{y'} \rtext{,} \\
  \Phi_{z} &= \gamma \, \Phi_{z'} \rtext{,} \\
  \Phi_{t} &= -v\gamma \, \Phi_{z'} \rtext{,}
\end{align}
which has already been incorporated into Eqs.~\eqref{eq:GEOT}--\eqref{eq:GEOZ}.
In order to numerically obtain a solution we need to specify boundary
conditions for the null geodesic. At the point of
emission of the light we demand
\begin{align}
  t &= 0 \rtext{,} 
 \\
  z &= z\sub s\rtext{,} \\
  y &= y\sub s\rtext{,} \\
  \dot{y} &= 0\rtext{,}
\end{align}
where $z\sub s$ and $y\sub s$ give the position of the source and have
been set to $-10\,000$ and $150$, respectively, in our calculations.
At the observer's position we impose the boundary condition
\begin{align}
  z &= z\sub o
\end{align}
with the position of the observer being set to $z\sub o=10\,000$ in our
calculation. Note that the $y$ position of the observer depends on the
deflection of the light ray and cannot be freely specified.

We solve this two-point-boundary value problem with a second order
accurate relaxation scheme. For our calculations we have set the mass
of the lens to $M=1$, so that all distances are given in units of the
lens' mass (remember that $c=1=G$).
These parameters lead to deflection angles much larger
than those observed in astronomical lenses.
We therefore expect this test of the \fod\ approximation to provide
upper bounds on the error and the \fod\ approximation to be
even more accurate in all scenarios of practical interest.

\section{Index of refraction\label{sec:refrindex}}

In this section we will provide an alternative descriptive
explanation of the
counterintuitive result obtained in Secs.~\ref{sec:obs frame}
and \ref{sec:lens frame} that a gravitational lens
moving towards the observer at highly relativistic speed leads to
a vanishing deflection angle. We have not found
an intuitive explanation for this effect
based on the components of the energy momentum tensor which diverge in the
limit of $v \rightarrow 1$, but whose contributions must cancel to
  explain the vanishing total effect.
Instead we describe the deflection in terms of
light moving in a refractive medium.

It is a well known fact that the deflection of light in weak \emph{static}
gravitational fields can be described by the analogous scenario of light
moving in Euclidean space filled with a refractive medium.
We therefore consider a refractive medium in the rest frame of the
lens. We denote the propagation speed of light in the medium by
$c'$ so that the refractive index is given by
\begin{align}
  n' &= \frac{1}{c'} = \frac{\diff t'}{\diff l'} \rtext{.}
        \label{eq:n'}
\end{align}
As before we use primed coordinates for the rest frame of the lens. Note,
however, that the spatial geometry is viewed as
Euclidian now, so that $\diff l'^2=\diff y'^2+\diff z'^2$.
The gravitational effect on the light propagation
is absorbed in the refractive index
\begin{align}
  n'-1 &= -2\Phi \rtext{,}
          \label{eq:n'2}
\end{align}
which follows from the combination of Eq.~\eqref{eq:n'} and
the line element \eqref{eq:metric static} for the case of light
(i.e., $\diff s^2=0$).
This description can be extended to the nonstatic scenario in the
rest frame of the observer either by applying the Lorentz
transformation \eqref{eq:lotra i}--\eqref{eq:lotra iii}
or by directly calculating $n$ as before but now using
the unprimed analogue of Eq.~\eqref{eq:n'} and the
metric \eqref{eq:metric moving} for the moving lens.
Both approaches lead
to the rather simple scaling law
\begin{align}
  \frac{n-1}{n'-1} &= \gamma^2 (1-v\cos\theta)^2 \rtext{.}
         \label{eq:n n'}
\end{align}
Here $\theta$ defines the direction of light propagation relative to the lens'
motion as seen by the
observer, i.e., $\cos\theta=\diff z/\diff l$, $\sin\theta=\diff y/\diff l$.
This result is
the generalization of the \fov\ expression of \citet{sef}.
It is also valid for arbitrary directions $\theta$, even though
we are still working in the \fod\ approximation.

We emphasize that we have obtained a description of a moving gravitational
potential in terms of
a refractive medium at rest but
with an anisotropic refractive index $n$ given by Eqs.~\eqref{eq:n'2}
and \eqref{eq:n n'}.
In particular this scaling law demonstrates that
the refractive index $n$ approaches unity in the limit of the gravitational
lens moving towards the observer with highly relativistic velocity
($\theta=0$, $v\rightarrow1$).
Translated back to the deflection of light, however,
a constant refractive index $n\equiv1$ implies a vanishing
of the deflection angle.

It is now not surprising at all that the refractive index
becomes unity in this limit.
The coordinate velocity of light is
reduced \emph{in the rest frame of the lens,} but
as the lens itself is moving highly relativistically,
the resulting net velocity \emph{in the observer's frame} is obtained
from relativistic velocity addition [cf.\ Eq.~\eqref{eq:vel comp}]
and approaches $1$. In the limit $v\rightarrow 1$ we thus
have no reduction of the velocity of light in the refractive medium and
consequently no deflection.

\section{Frame dragging\label{sec:framedrag}}

The fact that the velocity of light is reduced \emph{in a moving frame of
  reference} (cf.\ Appendix~\ref{sec:refrindex})
suggests effects analogous to the ``dragging of inertial frames'' in the field
of a rotating body with its extreme consequences close to Kerr black holes.
A discussion of this effect helps in understanding the geometric effects
in the field of a linearly moving mass and is therefore included in this
appendix. \citet{gabriel88} showed that the gravitomagnetic light deflection
of rotating bodies  can be expressed as an integral over the Lense-Thirring
rotation rate which demonstrates
the close relation between both effects.

To simplify the discussion we consider regions with a stationary (static
in the lens frame)
gravitational field in $y$ direction, i.e., $\Phi_z=\Phi_t=0$. It can be shown
easily that a freely falling particle will gain a velocity component in the
$z$ direction (parallel to the direction of motion of the lens) as a result
of its 
motion in the $y$ direction. This transversal acceleration is similar to
magnetic Lorentz 
forces and shows the gravitomagnetic effects directly.
Such effects are, however, coordinate dependent and thus not easy to
interpret in a general way. We therefore want to discuss effects on a body
\emph{at 
rest with respect to the outside world,} i.e., with constant $y$ and $z$
coordinates.

The tetrad of a local Minkowski coordinate system can be derived by starting
at one point of the world line and transforming the direction vectors $T^\nu$
with the Fermi-Walker transport which is defined by a vanishing Fermi
derivative,
\begin{align}
\frac{\Diff T^\alpha}{\Diff \tau} + g_{\mu\nu} \left(  u^\alpha
  \frac{\Diff u^\mu}{\Diff\tau} - u^\mu
  \frac{\Diff u^\alpha}{\Diff\tau} \right) T^\nu = 0 \rtext{.}
\label{eq:fermi}
\end{align}
Here the four-velocity of the test particle is denoted by
$u^\alpha=\diff
x^\alpha/\diff\tau$ (with $\tau$ being the proper time along the trajectory
of the test particle) and the covariant derivatives by an uppercase $\Diff$:
\begin{align}
\frac{\Diff T^\alpha}{\Diff \tau} &= \frac{\diff T^\alpha}{\diff \tau} +
\Gamma^\alpha_{\mu\nu} T^\mu u^\nu
\label{eq:covder}
\end{align}
In order to
derive the Christoffel symbols we write the Euler-Lagrange equations
[e.g., Eq.~\eqref{eq:elg y}]
for the Lagrange function in Eq.\,\eqref{eq:lagrange}.
On the other hand we write
the equations of motion by setting the covariant derivative of the
velocity vector $T^\alpha=u^\alpha$ in Eq.\,\eqref{eq:covder} to zero.
Even though we still work in the \fod\ approximation, we now do
allow for arbitrary directions of test particle velocities.
By comparing coefficients, we find the following nonvanishing
Christoffel symbols:
\begin{align}
\Gamma^t_{ty} &= (1+v^2)\gamma^2\Phi_y \ritext{,} & \Gamma^t_{yz} &= -2v\gamma^2\Phi_y \ritext{,}
 \\[1ex]
\Gamma^y_{tt} &= (1+v^2)\gamma^2\Phi_y \ritext{,} & \Gamma^y_{yy} &= -\Phi_y \ritext{,} \notag \\
\Gamma^y_{zz} &= (1+v^2)\gamma^2\Phi_y  \ritext{,}&
\Gamma^y_{tz} &= -2v\gamma^2\Phi_y  \ritext{,}\\[1ex]
\Gamma^z_{ty} &= 2v\gamma^2\Phi_y  \ritext{,}& \Gamma^z_{yz} &= -(1+v^2)\gamma^2\Phi_y \ritext{,}
\end{align}
as well as those following from the symmetry properties.
The four-velocity $u^\alpha$ of a particle at rest
is needed only to zeroth order of the deflection (i.e., potential)
and is given by $u^t=1$, $u^z=u^y=0$ in unprimed coordinates.
For the covariant derivative of $u^\alpha$ for the particle at rest, we
obtain in \fod\ approximation
\begin{align}
\frac{\Diff u^y}{\Diff \tau} = (1+v^2)\gamma^2\Phi_y
\end{align}
while all other components are vanishing.
This covariant derivative for the particle at rest is equivalent to the
  force needed to support the particle and prevent
  it from falling towards the  lens.

In the next step we apply
Eqs.\,\eqref{eq:fermi} and \eqref{eq:covder} to derive the
transformation of the components of an arbitrary vector $T^\alpha$. For
the metric $g_{\mu\nu}$, the zeroth order approximation
(i.e., the Minkowski metric)
is sufficient in this case. We arrive at the final equations for the
Fermi-Walker transport:
\begin{align}
\frac{\diff T^\alpha}{\diff \tau} &=
M^\alpha{}_\beta T^\beta \\
(M^\alpha{}_\beta) &= 
2v\gamma^2\Phi_y \begin{pmatrix} 0 & 0 & 0 \\ 0 & 0 & 1 \\ 0 & -1 & 0
  \end{pmatrix} \rtext{.}
\end{align}
The ordering of indices is $t,y,z$, the row index is $\alpha$, and the column
index $\beta$.
We note that the velocity vector $u^\alpha$ itself remains
constant as required. Furthermore, we see that the $y$ and $z$
coordinates evolve in a way which is equivalent to a rotation. Modulo an
arbitrary time offset, the fundamental solution of this part is
\begin{align}
T^y&=\cos(\omega \tau) \rtext{,} & T^z&=\sin(\omega \tau) 
\end{align}
with an angular velocity of $\omega=-2v\gamma^2\Phi_y$. As a consequence a
gyroscope (or any other device capable of sustaining a locally nonrotating
reference system) will precess relative to the outside world with this
angular velocity. This precession
can be observed from great distance
and represents a coordinate independent measure of
the frame dragging caused by the linear motion of the lens.\footnote{The
  assumption of constant spatial coordinates for the test particle is not
  coordinate dependent because it can be defined in the way that the
  test particle
  appears to be at rest when observed by a distant observer
  whose coordinates are
  fixed by the background Minkowski metric.}

Without presenting the calculations we note that the precession of a
gyroscope could be obtained more directly by solving the equation of motion
for the different parts of the rotating mass. The mass elements approaching
the lens will be dragged with the lens' motion while elements moving
parallel to the lens will feel a reduced gravitational attraction. It is
easy to show that this leads exactly to the precession which we derived in a
more formal but general way.

The effect is analogous to Lense-Thirring precession in the field of
rotating bodies, especially in the field of a Kerr black hole.
There is, however, one important difference. In our case,
all gravitomagnetic terms of the metric vanish in the rest frame of the
lens.
The precession due to the frame dragging
does of course not disappear when viewed in this system.
It would instead be interpreted as an incarnation
of geodesic precession of a \emph{moving}
particle in a \emph{static} gravitational field.
In the case of the Kerr metric, such a transformation
is possible locally only
as has been shown by \citet{ashby90}.

\def\araa{Annu.\ Rev.\ Astron.\ Astrophys.}
\def\annphys{Ann.\ Phys.\ (Leipzig)}
\def\mnras{Mon.\ Not.\ R.\ Astron.\ Soc.}
\def\prl{Phys.\ Rev.\ Lett.}
\def\aap{Astron.\ Astrophys.}
\def\pla{Phys.\ Lett.\ A}
\def\amjp{Am.\ J.\ Phys.}
\def\rpp{Rep.\ Prog.\ Phys.}
\def\phtrans{Philos. Trans. R. Soc. London}
\def\obs{The Observatory}
\def\nciml{Lett.\ Nuovo Cimento Soc.\ Ital.\ Fis.}


\end{document}